\documentclass[12pt]{article}
\usepackage{a4}
\usepackage{epsf}
\usepackage{amssymb}
\usepackage{cite}
\newcommand{\be}{\begin{equation}}
\newcommand{\ee}{\end{equation}}
\newcommand{\bea}{\begin{eqnarray}}
\newcommand{\eea}{\end{eqnarray}}
\newcommand{\pt}{\partial}

\begin{document}
\begin{titlepage}
\title{A Comment on Curvature Effects In CFTs And The Cardy-Verlinde Formula}
\author{}
\date{
Arshad Momen
\thanks{E--mail:~ momena@ictp.trieste.it, arshad@transbd.net, arshadmomen@yahoo.com}
\footnote{Permanent Address: Department of Physics, 
University of Dhaka, Dhaka-1000, Bangladesh}~~and~~Tapobrata Sarkar
\thanks{E--mail:~ tapo@ictp.trieste.it}
\vskip0.4cm
{\sl the Abdus Salam \\
International Center for Theoretical Physics,\\
Strada Costiera, 11 -- 34014~Trieste, Italy}}
\maketitle
\abstract{We examine the Cardy-Verlinde formula for finite temperature
$N=4$ Super Yang-Mills theory on $R\times S^3$, and its AdS dual. We find
that curvature effects introduce non-trivial corrections to thermodynamic
quantities computed on both sides. We find a modified version of the 
Cardy-Verlinde formula for the SYM theory, incorporating these. On the gravity 
side, these corrections imply that the Cardy-Verlinde
formula is exact.}
\end{titlepage}

The Holographic principle \cite{hol} implies that the degrees of freedom in
a given spatial volume can be thought of as being encoded on its boundary, with 
a density that does not exceed one degree of freedom per Planck area.  
The basic idea behind this principle is the fact that the maximum possible entropy 
that one can associate with the given volume is that of the largest
black hole that can be fitted inside this volume. The Bekenstein-Hawking entropy
of the black hole then gives rise to the the constraint
\be
S \leq \frac{A}{4G}
\label{i1.1}
\ee
where $A$ is the horizon area of the maximal black hole, and $G$ is Newton's 
constant. Recently, a lot of progress has been made in the understanding of the 
Holographic principle, both in the context of string theory and that of canonical
gravity. In fact, many of these recent attempts rely on techniques of conformal 
field theories in two dimensions. The Cardy formula, that gives a
prescription for counting states in a two dimensional conformal theory is at the
centre of these developments. Using the AdS/CFT duality \cite{adscft}, 
E. Verlinde \cite{verlinde} has proposed a generalisation of the 
Cardy formula (now commonly referred to as the Cardy-Verlinde formula), valid 
for arbitrary space-time dimensions, which count the density of states 
for CFT's that have an AdS dual. 

It is now known that for the Cardy formula for two
dimensional  conformal field theories there are logarithmic corrections
\cite{carlip}. Similar corrections which are
logarithmic has also been discussed in the context of black
hole thermodynamics  \cite{DMB} originating  from thermal fluctuations in
the system.  Therefore, it is not unnatural to anticipate corrections to the 
holographic bound (\ref{i1.1}) or to the Cardy-Verlinde formula.

However, there could be other non-trivial subleading corrections to the
holographic bound. Consider type IIB string theory
on $AdS_5 \times S^5$, which is dual to $N=4$ $SU(N_c)$ Super
Yang-Mills (SYM) theory on  $S^3 \times R$. 
The gravity theory  on  $AdS_5$  admits blackhole
solutions which correspond to the $N=4$  SYM theory at high temperatures
(thermalization being achieved in the usual way of compactifying the
Euclidean time direction on a circle, and identifying the temperature as
the inverse of the circle radius). The Cardy-Verlinde formula seems to be
satisfied in this picture \cite{verlinde}. Now, one recalls that
when the space is flat, $N=4$ SYM is superconformal. Thus, from the
point of conformal theories, one is tempted to consider curvature perturbations. 

At first sight, this might seem to be a non-issue, since the high-temperature
limit assumes, in a sense to be made more precise later, that the curvature
is large. However, one could still try to estimate the order of
magnitude corrections (if any) even within these limitations. If, for example,
one is able to find such corrections from purely curvature effects on the 
CFT side, this could mean a correction to the Bekenstein bound in the theory.

It is this issue that we study in this note. Specifically, we study the
finite temperature $SU(N_c)$ SYM theory in a curved background, and calculate the
various thermodynamic quantities. We find that indeed there are subleading 
order corrections to the energy and entropy for such a theory in a curved 
background at high temperatures that leads to a certain modification of the
Cardy-Verlinde formula.
 
This note is organised as follows. First, we review the basic CFT 
framework that we are going to work in, and proceed to compute leading order
corrections to the various thermodynamic quantities and an analogue of the
Cardy-Verlinde formula. Next, we compute similar quantities on the 
gravity side, and compare the results. We end with some discussions and 
conclusions. 

\vspace{1.0cm}

Consider an observer in a universe with flat spatial section, who carries out 
thermodynamic measurements on a system
consisting of a massless gas at some given temperature $T$.  
We will hereafter view the flat space as the infinite radius (hence zero 
curvature limit of $S^3$). Let the volume of the gas be $V$. Standard 
finite temperature calculations lead to a free energy density
\be
{ \cal F} =- \frac{\pi^2 N_c^2 T^4 }{6} .
\label{s1.0}
\ee

The presence of the fourth power of temperature $T$ in the above expression
leads to the tracelessless of the energy-momentum tensor, as in the case of
blackbody radiation.

Now, consider turning on  a small but nonzero constant curvature to
the space, characterised by the radius of curvature  $l$. We wish to study
the thermodynamics of the $N=4$ SYM gas in this space. The free energy
density for such curved backgrounds has been considered by
Burgess, Constable and 
Myers \cite{BCM}.  In the conformal limit, this system has an AdS dual \cite{adscft}.
Hence, by comparing with the dual theory in this limit, we would be able to derive
the Cardy-Verlinde formula for this system, as well as identify curvature corrections.

We start with the free energy density of the four dimensional N=4 Super 
Yang-Mills on an Einstein Manifold at a finite temperature which 
has been calculated in \cite{BCM}, 
using heat kernel techniques - 
\be
{ \cal F} =- \frac{\pi^2 N_c^2 }{6} \left[T^4 + \kappa \frac{3 T^2}{2 
\pi^2 l^2} + \cdots \right].
\label{s1.1}
\ee
   Here the spatial geometry for the system is taken to be a maximally 
symmetric three dimensional space so that the Ricci scalar on this space 
is given by $R = -\frac{6 \kappa}{l^2}$, $l$ being the ``radius'' of the space.
When $\kappa =+1, 0, -1$, the space is given by a three sphere, flat 
space or hyperbolic space respectively.Hereafter we will exclusively deal
with a closed space ,i.e, a sphere  and hence $\kappa = 1 $. Now note that 
when $l \rightarrow \infty$ one recovers the standard flat space result. 
Therefore the second term in the above free energy density (which is 
proportional to $T^2$) can be thought of as the contribution from the 
background geometry to the thermodynamic free energy of the system. 
The terms represented by the ellipsis are of the order $O( \frac{1}{T^2 l^6})$. 
Also, note that in the case of the sphere geometry, the volume of the sample
is bounded from above, $V\leq 2\pi^2l^3$. 
In \cite{BCM}, the quantity ${\cal F}(T)-{\cal F}(0)$ was calculated, and
so, there can be a temperature independent term, as calculated
by Kutasov and Larsen \cite{kutlar}. However, for the moment we supress 
this term. We will come back to this issue in a while.  
  
Recall that the thermodynamic measurements on the $N=4$ SYM gas will require the 
volume of the system $V$ to be a variable. Thus,  the free energy for
the  finite volume system is given by
\be
F (T,V) = {\cal F} V = -\frac{\pi^2 N_c^2 V }{6} \left[T^4 - \kappa \frac{3 T^2}{2 \pi^2 l^2} + \cdots \right].
\label{s1.2}
\ee
For the moment, we keep the volume of the 3-sphere, $V$, distinct from the radius
of curvature, $l$ for convenience.
 
Given the Helmholtz free energy one can readily find the entropy of the 
gas 
\be
S = -\frac{ \pt F}{\pt T} = \frac{ 2 \pi^2 N_c^2 V}{3} \left[ T^3 - 3\frac{T}
{4\pi^2 l^2} + \cdots \right].
\label{s1.3}
\ee

Thus the internal energy of the ``gas'' would be given by 
\be 
E = F + TS = \frac{\pi^2 N_c^2 V}{2}P \left[T^4 - \frac{T^2}{2 \pi^2 l^2 } + 
\cdots \right].
\label{s1.4}
\ee
The ellipses in the above equations represent terms of order
$O(\frac{1}{l^6})$. Since, the internal energy of the system is intrinsically 
a function of the entropy and the volume, we have to reexpress $E$ in terms of 
$S$ and $V$. This can be done via the elimination of $T$ using equation 
(\ref{s1.3}). For high temperatures, this equation is a cubic equation 
in $T$:
\be
T^3 - 3 \frac{T}{4 \pi^2 l^2 } - \frac{3S}{2 \pi^2 N_c^2 V} = 0,
\label{s1.5}
\ee   
and hence can be solved by elementary Cardano's formula (the temperature 
independent term in (\ref{s1.1}) does not appear here). Now, in the limit 
$\alpha \equiv \frac{N_c^2 V}{6 \pi S l^3} < 1$, the equation has a pair 
of complex conjugate roots and a real root. Note that for the sphere 
$ V \leq 2 \pi^2 l^3 $ and the above condition gives a further restriction
$ S > \frac{\pi N_c^2 }{3} $. Now, we work in the limit $ \alpha << 1$, 
in which the temperature of the gas is given by 
\be
T = T_0\left( 1 + \frac{1}{(2\pi l  T_0)^2 }\right) \qquad \qquad T_0 \equiv 
\left[\frac{3S}{2\pi^2 N_c^2 V}\right]^{\frac{1}{3}},
\label{s1.6}
\ee
$T_0$ will now have the interpretation of the temperature of the gas in the 
limit $l\rightarrow\infty$. This shows that gas 
becomes ``hotter'' in the presence of (negative) curvature. 

Let us now spell out exactly the limits in which we work, 
in terms of the several quantities that have been defined. Firstly
the original expression (\ref{s1.1}), is valid for large
$N_c$, and in the high-temperature limit, i.e $\frac{1}{lT}<<1$.
We will assume, further, that we are working in the large curvature limit,
and hence we will choose both $l$ and $T$ to be large. Then, it is 
clear from (\ref{s1.5}) that we are dealing with large entropies.

Substituting (\ref{s1.6}) in the expression for  the internal energy
one gets $E$ as a function of entropy and the volume $V$ of the gas -
\be
E = \frac{\pi^2 N_c^2 }{2 } \left[ 
\left(\frac{3S}{2 \pi^2 N_c^2}\right)^{\frac{4}{3}} V^{- \frac{1}{3}} +   
\frac{1}{2 \pi^2 l^2}\left( \frac{3S}{2 \pi^2 N_c^2}\right)^{\frac{2}{3}} 
V^{\frac{1}{3}} + \frac{V}{8 \pi^4 l^4 } \right]
\label{s1.7}
\ee
Let us analyse this equation more carefully. 
When one substitutes $V=2\pi^2l^3$, as is appropriate for the three-sphere,
we do get three contributions. The first term now scales as $S^{4/3}/l$,
and the second term as $S^{2/3}/l$. These are the known behaviours for the
``extensive'' and ``Casimir'' part of the energy \cite{verlinde}. However,
now a finite contribution to internal energy has appeared 
which can be thought of as a gravitational contribution to the internal 
energy from the spatial curvature, which scales as $1/l$. Note that, to 
begin with, we neglected the temperature independent contribution to the
free energy. Clearly, the last term in (\ref{s1.7}) is of this form. 
In fact, its effect is to lower the magnitude of the temperature-independent
term that arises in the free energy as computed in \cite{kutlar}. Interestingly,
in the high-temperature limit, the $T$-independent term will presicely 
cancel out, as we will see later. 

It is further tempting to note that 
if we calculate the Casimir energy for the system, using the fact that
it is essentially the violation of the Euler identity,
\be
E_C = 3\left(E+pV-TS\right)
\label{s1.8}
\ee
we get $E_C=0$ if the volume $V$ is treated independent of the radius of
curvature. In other words,  from eq. (\ref{s1.7}), it is obvious that
once we have treated $l$ and $V$ at a different footing ($V$ involving a
length scale and hence driving the system away from conformality), the energy scales
as $E(\lambda S,\lambda V)=\lambda E(S,V)$. Therefore there is no violation
of the Euler identity. However, for this situation, the formula for the 
free energy density will be different than the one in \cite{BCM}, when 
one does the full QFT calculation. It will be interesting to study the 
Cardy-Verlinde formula for such non-conformal cases, i.e when the volume in 
which the observer carries out his measurements introduces another length scale 
in the problem.

Once we identify the 
volume $V$ with the maximal volume of the three-sphere, i.e $2\pi^2l^3$, we
get the following relation for the entropy
\be
E'=\frac{\pi^2N_c^2}{2}\left[\left(\frac{3S}{2\pi^2N_c^2}\right)^{4\over 3}
V^{-{1\over 3}}+\frac{1}{(2\pi^2)^{1\over 3}}\left(\frac{3S}{2\pi^2N_c^2}\right)
^{2\over 3}V^{-{1\over 3}}\right]
\label{s1.15}
\ee
The l.h.s of the above equation representing a ``shifted'' energy, i.e, the
energy $E$ shifted by temperature-independent terms, proportional to $V$.
The entropy, from eq. (\ref{s1.7}), can be written as
\be
S={\sqrt 2}\frac{2\pi l}{3}\sqrt{2E_EE_C} 
\label{s1.16}
\ee
Note that we have a prefactor of $\sqrt 2$ which is different from the 
factor of $2$ obtained in \cite{klemm}. 
There is an upper bound on $S$ in eq.(\ref{s1.16}), namely,
\be
S\leq{\sqrt 2}\frac{2\pi}{3}E'l
\label{s1.17}
\ee
where again $E'$ is the shifted energy. The temperature independent
contribution to the free energy is \cite{kutlar} $\frac{-3}{16}\frac{N_c^2}{l}$. 
Along with a similar contribution that arose in (\ref{s1.7}), 
which modifies this term, we obtain finally,
\be
S\leq{\sqrt 2}\frac{2\pi}{3}El + \frac{N_c^2V}{32\pi^2l^3}
\label{s1.18}
\ee
Of course, when $l$ is very large, the standard Bekenstein bound
is reproduced, although with a different prefactor than \cite{verlinde}. 
\footnote {From a completely independent perspective \cite{carlip} there
are the so called ``logarithmic corrections'' to entropy in a conformal theory
which reduces the upper bound on the entropy.} 

\vspace{1.0cm}
The AdS/CFT correspondence of Maldacena \cite{adscft} 
relates the strongly coupled four dimensional ${\cal N}=4$ SYM to 
$AdS_5$ supergravity. One can try to check the 
validity of the above results in the context of the AdS/CFT 
correspondence. We do this again by looking at the free energy density
of the strongly coupled system, which is obtained by looking at the 
black hole solution of the 5D AdS gravity, and using the AdS/CFT
correspondence \cite{BCM}:  
\be
{\cal F} = -\frac{\pi^2 N_c^2 }{8} \left[ T^4 - \frac{3 T^2 }{4 \pi^2 l^2} + 
\frac{1}{8 \pi^4 l^4} \right]
\label{s2.1}
\ee 
In the same way as before, solving for the temperature in terms of the 
entropy and feeding it back into the expression for the energy, we
obtain the equation
\be
E=\frac{3}{8}\pi^2N_c^2V\left[\left(\frac{2S}{\pi^2N_c^2V}\right)^
\frac{4}{3}+\frac{1}{\pi^2l^2}\left(\frac{2S}{\pi^2N_c^2V}\right)^
\frac{2}{3}\right]
\ee
Notice that here, in contrast with the previous case, the temperature-independent 
term in eq. (\ref{s2.1}) exactly cancels with the curvature term coming from 
the high temperature expansion of $T$. Hence, the energy has an extensive and
a sub-extensive term only. It is easily shown that in this case, 
the Cardy-Verlinde formula is
\be 
S=\frac{2\pi l}{3}\sqrt{2E_EE_C}
\ee
with $E_E$ and $E_C$ being the extensive and sub-extensive parts of the
energy respectively. Hence, the Cardy-Verlinde formula, in this regime, is
exact. 

\vspace{1.0cm}
In summary, we have taken a critical look at the Cardy-Verlinde formula 
\cite{verlinde} for ${\cal N}=4$ $SU(N_c)$ SYM at high temperatures and
its AdS dual theory. Our tools have been elementary, but using these,
we have been able to take into account systematically the curvature
and high temperature effects in the relations defining the thermodynamics
of the system. Our main result in this note is that there is a universal
Cardy-Verlinde formula, which is exact at strong coupling. The high-temperature
expansion of the thermodynamic quantities ensured in that case that the
temperature-independent terms indeed drop out of the expression for the 
energy, and also the Cardy-Verlinde formula holds exactly. In the weak 
coupling regime, this formula is true upto a factor of $\sqrt 2$, this is
in contrast with the result obtained by Klemm et. al \cite{klemm}, differing
from their result by a further factor of $\sqrt 2$. 

An interesting possibility that one might consider for future
work is to investigate if an analogue of the Cardy-Verlinde formula holds for 
a generic system that is perturbed away from the conformal limit. It will
also be interesting to consider the effects of logarithmic corrections to
the various well known holographic bounds. We leave such considerations for 
a future publication. 

\vspace{1.0cm}
\noindent
{\bf Acknowledgements}

\noindent
We would like to thank D. Klemm for helpful correspondence. 
We would also like to thank M. Blau, K.S. Narain, G. Thompson  and  M. 
Zamaklar for discussions. A. M. would like to thank Prof. S. 
Randjbar-Daemi and the other members of the HE group for the warm  
hospitality and support at ASICTP where this work was done.

\end{document}